\documentclass[10pt,preprint]{aastex}
\usepackage{epsfig}
\newcommand{\myemail}{rilett@physics.montana.edu}

\begin{document}

\title{MHD Shock Conditions for Accreting Plasma onto Kerr Black Holes - I}

\author{Masaaki Takahashi}
\affil{Department of Physics and Astronomy, Aichi University of
       Education, Kariya, Aichi 448-8542, Japan;
       takahasi@phyas.aichi-edu.ac.jp}
\and
\author{Darrell Rilett,  Keigo Fukumura,  and  Sachiko Tsuruta}
\affil{Department of Physics, Montana State University,
       Bozeman, MT 59717-3840;
       \myemail, fukumura@physics.montana.edu, tsuruta@physics.montana.edu}

\begin{abstract}
We extend the work by Appl and Camenzind (1988) for special relativistic  
magnetohydrodynamic (MHD) jets, to fully general relativistic studies  
of the standing shock formation for accreting MHD plasma in a rotating,  
stationary and axisymmetric black hole magnetosphere.  All the postshock  
physical quantities are expressed in terms of the relativistic 
compression ratio, which can be obtained in terms of preshock 
quantities. Then, the downstream state of a shocked plasma is determined 
by the upstream state of the accreting plasma. In this paper sample 
solutions are presented for slow magnetosonic shocks for accreting flows 
in the equatorial plane. We find that some properties of the slow 
magnetosonic shock for the rotating magnetosphere can behave like a fast 
magnetosonic shock. In fact, it is confirmed that in the limit of weak 
gravity for the upstream non-rotating accretion plasma where the 
magnetic field lines are leading and rotating, our results are very 
similar to the fast magnetosonic shock solution by Appl and Camenzind 
(1988). However, we find that the situation becomes far more 
complicated due to the effects of strong gravity and rotation, such as 
the frame dragging-effects. We show the tendency that the large spin of 
the black hole makes the slow magnetosonic shock strong for the 
accretion solutions with the same energy-flux.   
\end{abstract}


\section{Introduction}

Recent theoretical and observational developments suggest that various 
members of active galactic nuclei (AGN) involve supermassive black holes 
(e.g., \cite{rees97}). According to the `standard unified scenario' an 
AGN reveals itself as a quasar or a Seyfert nucleus during the earlier 
stages when the accretion rate is high and the black hole is spinning 
up, while in later stages when the accretion slows down and the black 
hole starts losing its rotational energy by spinning down it evolves to 
a radio galaxy.  In the past years physics of black hole magnetospheres 
has been extensively studied, mainly in connection with winds, jets, 
and energy extraction from black holes in rotation-powered AGN 
(\cite{bz77,zk77,tpm86}; see, e.g., \cite{bbr-rev} for comprehensive 
review). An important step for studying the magnetosphere in the Kerr 
background space-time was originated by \cite{bz77} in the force-free 
limit. The next logical step will be to extend these magnetospheric 
studies to accretion-powered AGN --- especially to Seyfert nuclei 
where the accretion rate should be relatively moderate. 
\cite{ph83} first extended his general studies of black hole 
magnetospheres to accretion-powered AGN, adopting the 
magnetohydrodynamics (MHD). 
In the early 1990s, \citet[hereafter TNTT90]{tk90}, \cite{ni91} and 
\cite{hi92} investigated both inflow and outflow of material in black 
hole magnetospheres. Their approach was a general relativistic unified 
MHD description of both matter accretion process and electromagnetic 
process in the magnetosphere. Relaxing the force-free limit, these 
authors explored the roles of interactions between the accreting fluid 
matter and electromagnetic fields.

Recently, ASCA observations indicated evidence that the Fe lines 
observed from some Seyfert nuclei are emitted from regions very 
close to the central black hole (e.g., \cite{tn95,nd97}). \cite{iw96} 
interpreted the behavior of Seyfert MCG 6-30-15 deduced from the long 
ASCA observation as an indication that the black hole is rotating 
extremely fast, near the maximum limit. \cite{ry97} pointed out that 
such extremely fast rotation contradicts with the standard unified 
scenario. Instead, these authors showed that if an X-ray point source 
somewhere above the hole on the rotation axis illuminates the infalling 
gas within the inner accretion disk radius, a Schwarzschild hole is 
consistent with the observation.  However, it is not clear how such 
an X-ray source can be created in the specified location. This issue 
clearly points to {\it the importance of investigating, for 
accretion-powered AGN also, the basic physics of the vicinity very 
close to a black hole} --- especially the regions between the inner 
boundary of the accretion disk and the event horizon.  As our first 
step toward such investigation, therefore, we explored the standing 
shock formation in accreting plasmas in black hole magnetospheres.  
These studies may shed valuable insight to problems such as how X-ray 
sources can be created near the event horizon.

We consider the central engine of an AGN as a black hole magnetosphere, 
where magnetized plasma surrounds a black hole and infalling accreting 
flows and outgoing wind/jets would be generated from the surrounding 
plasma. We formulate the MHD shock conditions in Kerr geometry for 
such plasma.  The formation of shocks is based on the existence of 
multi-magnetosonic points in the accretion solution, while for the 
outgoing plasma the sub-magnetosonic solution is allowable at distant 
regions. This is because, for instance, the accreting flows initially 
ejected from a plasma source with low velocity must be terminally 
super-fast magnetosonic at the event horizon. At the shock front, the 
flow transits from super-magnetosonic to sub-magnetosonic, so that the 
accreting flows with a shock must pass through a magnetosonic point on 
each side of the shock front.
This situation is quite similar to the case of hydrodynamical 
accretion onto a black hole (e.g., \cite{ch90a,sm94,lu97,lu98}).
The trans-magnetosonic MHD flow solution was discussed by 
\cite{tk2000a,tk2000b,tk2001}. Along the magnetic field line, the five 
physical quantifies are conserved: the total energy $E$ and angular 
momentum $L$, the angular frequency of the magnetic field line 
$\Omega_F$, the particle number flux per magnetic flux tube $\eta$, and 
entropy $S$ (see, e.g., \cite{ca86}). When these conserved quantities 
are specified at the plasma source, the location of the fast/slow 
magnetosonic points and the Alfv\'en point are determined.  
\cite{tk2000a} obtained multi-magnetosonic point solutions and found 
two regimes of accretion flows -- ``hydro-like'' and ``magneto-like''. 
The hydro-like accretion would transit to magneto-like accretion by the 
shock formation.  However, we will postpone, until later, the problem 
of joining these two types of solutions by shock formation. The reason 
is that in order to do so we will have to carry out detailed parameter 
search for trans-magnetosonic MHD flows, but then we will have five 
field-aligned physical quantities in both upstream and downstream 
solutions, not a trivial situation. 
This problem will be additional in a subsequent paper (Rilett et al., 
in preparation).  This paper is meant as a starting 
point for our long-range investigation of shock conditions for accretion 
flows in the Kerr geometry.  Therefore, here we will only solve the cold 
trans-fast MHD equations for upstream accretion and discuss the shock 
properties at the shock fronts. At the shock front the plasma would be 
heated up. So, the postshock accretion should be treated by a hot MHD 
accretion model \citep{tk2001}. In this paper we will not yet solve 
explicitly the hot trans-magnetosonic solutions for postshock accretion 
but instead treat the shock front location as a free parameter. However, 
by joining preshock and postshock solutions, the shock location would be 
determined at one (or more) place(s).

The main purpose of this paper is to explore the effects of rotation 
and general relativity on the MHD shock conditions for accreting plasma 
in a black hole magnetosphere. In order to do so, first in Section 2 we 
present the basic equations for MHD accretion in the Kerr geometry. 
Then, in Section~3 we extend the work for special relativistic MHD 
jets by~\citet[hereafter AC88]{ac88}, by deriving the shock conditions 
for general relativistic MHD accretion onto a Kerr black hole. 
(Non-relativistic MHD outflow solutions with standing shocks were 
 discussed by \cite{ch90b}.) 
Following AC88, all of our postshock physical quantities are expressed 
in terms of the relativistic compression ratio $\xi$. This compression 
ratio is the solution of a polynomial of eighth degree.  Due to the 
additional factors, namely magnetosphere rotation and general relativistic 
effects, the mechanism of solution is far more complicated and tedious 
than in AC88. Our equations reduce to the AC88 equations in the weak 
gravity limit without plasma rotation. 
TNTT90 extensively studied MHD accretion flows in the Kerr geometry. 
As the next step our shock conditions are applied to the accreting 
MHD plasma flows as described by TNTT90. 
In Section~4 we present some examples of representative physically 
relevant shock solutions found for acceptable accretion flows onto 
the event horizon. Our results are presented for the equatorial 
flows and slow magnetosonic shocks. 
The ``switch-off'' shock (switching-off of the magnetic field at the 
shock front) and comparison between AC88 and our results are also  
discussed. More thorough presentation of the results and extensive 
discussion of the physical implication of these results will be given 
in Paper II (Rilett et al., in preparation). The general cases including 
the fast magnetosonic shocks will be presented in Paper III (Rilett 
et al., in preparation).  The possible application of our current work 
to some astrophysical problems, such as the Seyfert Fe lines, will be 
given in subsequent papers. Summary and conclusion are given in 
the last section 5.

\section{Basic Equations of General Relativistic Plasma Flow }

In this section we summarize the general relativistic MHD flows 
(see, e.g.,~\cite{ca86b,ca89}; TNTT90). We assume a stationary and 
axisymmetric magnetosphere and ignore its self-gravity. We also require 
infinite conductivity for the plasma flow. The background metric is 
given by the Boyer-Lindquist coordinates with the $c=G=1$ units  
\begin{eqnarray}
   ds^2 &=& \left( 1-\frac{2mr}{\Sigma} \right) dt^2
        + \frac{4amr\sin^2\theta}{\Sigma} \,dt d\phi \\
        & & - \frac{A\, \sin^2\theta}{\Sigma} \, d\phi^2
        - \frac{\Sigma}{\Delta}\, dr^2 - \Sigma\, d\theta^2  \ ,  \nonumber
\end{eqnarray}
where 
$\Delta \equiv r^2 -2mr +a^2 $, 
$\Sigma \equiv r^2+a^2\cos^2\theta$, 
$A \equiv (r^2+a^2)^2-a^2\Delta \sin^2\theta$ and  $m$ and $a$ 
denote the mass and angular momentum per unit mass of the black 
hole, respectively. The basic equations of relativistic plasma 
flows are as follows: (i) the particle number conservation 
\begin{equation}
  (nu^\alpha)_{;\alpha} = 0 \ ,
\end{equation}
where $n$ is the proper particle number density and $u^\alpha$ is the 
fluid 4-velocity; (ii) the conservation of total energy and momentum 
\begin{equation}
  T^{\alpha\beta}_{;\beta}=0 \ , \label{eq:basic-2} 
\end{equation}
where the energy-momentum tensor is given by 
\begin{equation}
  T^{\alpha\beta}=n\mu u^\alpha u^\beta -Pg^{\alpha\beta} 
  + \frac{1}{4\pi}\left(F^\alpha_{\ \lambda} F^{\lambda\beta}
  + \frac{1}{4}g^{\alpha\beta}F^2 \right) \ ,
\end{equation}
$\mu$ is the relativistic specific enthalpy, $P$ is the pressure, 
$F^{\alpha\beta}$ is the electromagnetic field tensor and 
$F^2=F^{\alpha\beta}F_{\alpha\beta}$; and (iii) the MHD condition 
\begin{equation}
  u^\beta F_{\alpha\beta}=0 \ .
\end{equation}

The magnetic field and electric field seen by a distant observer are 
defined as 
\begin{eqnarray}
  B_\alpha &\equiv&
  \frac{1}{2}\eta_{\alpha\beta\gamma\delta}k^\beta F^{\gamma\delta} \ , \\
  E_\alpha &\equiv& F_{\alpha\beta}k^\beta                            \ ,
\end{eqnarray}
where $k^\alpha=(1,0,0,0)$ is the time-like killing vector and 
$\eta_{\alpha\beta\gamma\delta}\equiv 
   \sqrt{-g}\, \epsilon_{\alpha\beta\gamma\delta}$.
Here, we can introduce the angular velocity of the magnetosphere 
(see,~\cite{bk78}) 
\begin{equation}
  \Omega_F \equiv -\frac{F_{tr}}{F_{\phi r}}
   = -\frac{F_{t\theta}}{F_{\phi\theta}} \ ,  
\end{equation}
which is conserved along magnetic field lines. 
Then, we obtain the following useful expressions 
\begin{eqnarray}
   E_\theta &=& -\sqrt{-g}\, \Omega_F B^r      / G_t  \ ,  \label{eq:Et} \\
   E_r      &=&  \sqrt{-g}\, \Omega_F B^\theta / G_t  \ ,  \label{eq:Er}
\end{eqnarray}
where $G_t\equiv g_{tt}+g_{t\phi}\Omega_F$.

From the conservation of total energy and momentum~(\ref{eq:basic-2}) 
and the Killing equation $\chi_{\mu;\nu}+\chi_{\nu;\mu}=0$, where 
$\chi^\mu$ is a Killing vector, it follows that 
$(\chi_\mu T^{\mu\nu})_{;\nu}=0$ (see,~\cite{bz77}). Thus, for any 
stationary and axisymmetric system, we define the conserved energy flux 
\begin{equation}
  {\cal E}^\mu = T^{\mu\nu}k_\nu = T^\mu_t
   = (T^\mu_t)_{\rm em}+(T^\mu_t)_{\rm fluid}
\end{equation}
and angular momentum flux 
\begin{equation}
  -{\cal L}^\mu = T^{\mu\nu}m_\nu = T^\mu_\phi
   = (T^\mu_\phi)_{\rm em}+(T^\mu_\phi)_{\rm fluid} \ ,
\end{equation}
where $m^\nu$ is the axial Killing vector with Boyer-Lindquist component 
(0,0,0,1), and the electromagnetic part and fluid part are labeled by 
``em'' and ``fluid'', respectively. The fluid and electromagnetic parts 
of the radial component of energy flux ${\cal E}^r$ are 
\begin{eqnarray}
  {\cal E}^r_{\rm fluid} &\equiv& (T^r_t)_{\rm fluid}
         \ = \ n\mu u^r u_t \ , \\
  {\cal E}^r_{\rm em} &\equiv& (T^r_t)_{\rm em}
         \ = \ -\frac{B_\phi E_\theta}{4\pi\sqrt{-g}}
         \ = \ \frac{B_\phi B^r \Omega_F }{4\pi G_t} \ ,
\end{eqnarray}
and the fluid and electromagnetic parts of the radial component of 
angular momentum flux ${\cal L}^r$ are 
\begin{eqnarray}
  -{\cal L}^r_{\rm fluid} &\equiv& (T^r_\phi)_{\rm fluid}
           \ = \ n\mu u^r u_\phi \ , \\
  -{\cal L}^r_{\rm em} &\equiv& (T^r_\phi)_{\rm em}
           \ = \ -\frac{B_\phi}{4\pi g_{tt}} \left(
  \frac{g_{t\phi} E_\theta}{\sqrt{-g}}+ B^r \right)
           \ = \ -\frac{B_\phi B^r}{4\pi G_t} \ ,
\end{eqnarray}
where we have used the relation~(\ref{eq:Et}).

We can express the radial energy and angular momentum fluxes as 
${\cal E}^r =nu^r E$ and ${\cal L}^r =nu^r L$, respectively, where 
$E$ and $L$ are the total energy and the total angular momentum of 
the MHD flow seen by a distant observer and defined by 
\begin{eqnarray}
  E &\equiv& \mu u_t - \frac{\Omega_F B_\phi}{4\pi\eta} \ , \\
  L &\equiv& -\mu u_\phi - \frac{B_\phi}{4\pi\eta} \ ,
\end{eqnarray}
and $\eta$ is the particle number flux per magnetic flux tube defined by 
\begin{equation}
  \eta \equiv -\frac{nu^r}{B^r}G_t \ . \ \ \label{eq:eta}
\end{equation}
These quantities $E$, $L$ and $\eta$ are conserved along stream 
lines, which coincide with magnetic field lines (see,~\cite{ca86}).

The energy and angular momentum of the fluid are 
\begin{eqnarray}
  \mu u_t &=& \frac{M^2E - e G_t}{M^2-\alpha}         \ \ , \label{eq:ut} \\
  \mu u_\phi &=&-\ \frac{M^2L + e G_\phi}{M^2-\alpha} \ \ , \label{eq:ufai}
\end{eqnarray}
where $M^2\equiv 4\pi\eta^2(\mu/n)$ is the Alfv\'en Mach number, 
$e\equiv E-\Omega_F L$, 
$G_\phi \equiv g_{t\phi}+g_{\phi\phi}\Omega_F$ and 
$\alpha\equiv G_t+G_\phi \Omega_F$.

The poloidal equation (the relativistic Bernoulli equation) 
of the magnetized flow is written as 
\begin{equation}
  (1+u_p^2) = (E/\mu)^2[(\alpha-2M^2)f^2-k] \ , \label{eq:pol-eq}
\end{equation}
where 
\begin{eqnarray}
  f &\equiv& -\frac{G_\phi+G_t \tilde L}{\rho_w(M^2-\alpha)} \ , \\
  k &\equiv& -(g^{tt}-2g^{t\phi}{\tilde L}+g^{\phi\phi}{\tilde L}^2) \ ,
\end{eqnarray}
$u_p$ is the poloidal velocity defined by 
$u_p^2\equiv -u^A u_A$ ($A=r, \theta$),  
$\rho_w^2 \equiv g_{t\phi}^2-g_{tt}g_{\phi\phi} = \Delta\sin^2\theta$ 
and $\tilde L \equiv L/E$. We set $u_p>0$ ($u^r<0$) for ingoing flows.
When we neglect dissipative effects and cooling in the plasma, 
the flow is adiabatic with a specific enthalpy of the form 
(see,~\cite{ca87}) 
\begin{equation}
  \mu = m_{\rm p}+\frac{\Gamma}{\Gamma-1}\frac{P}{n} \ , \label{eq:muP}
\end{equation}
where $m_{\rm p}$ is the rest mass of the particle. 
The toroidal component of the magnetic field can be expressed as 
\begin{equation}
  B_\phi = 4\pi\eta\frac{G_\phi E + G_t L}{M^2-\alpha} \ . \
\end{equation}

\begin{figure}
    \epsscale{0.7}
    \plotone{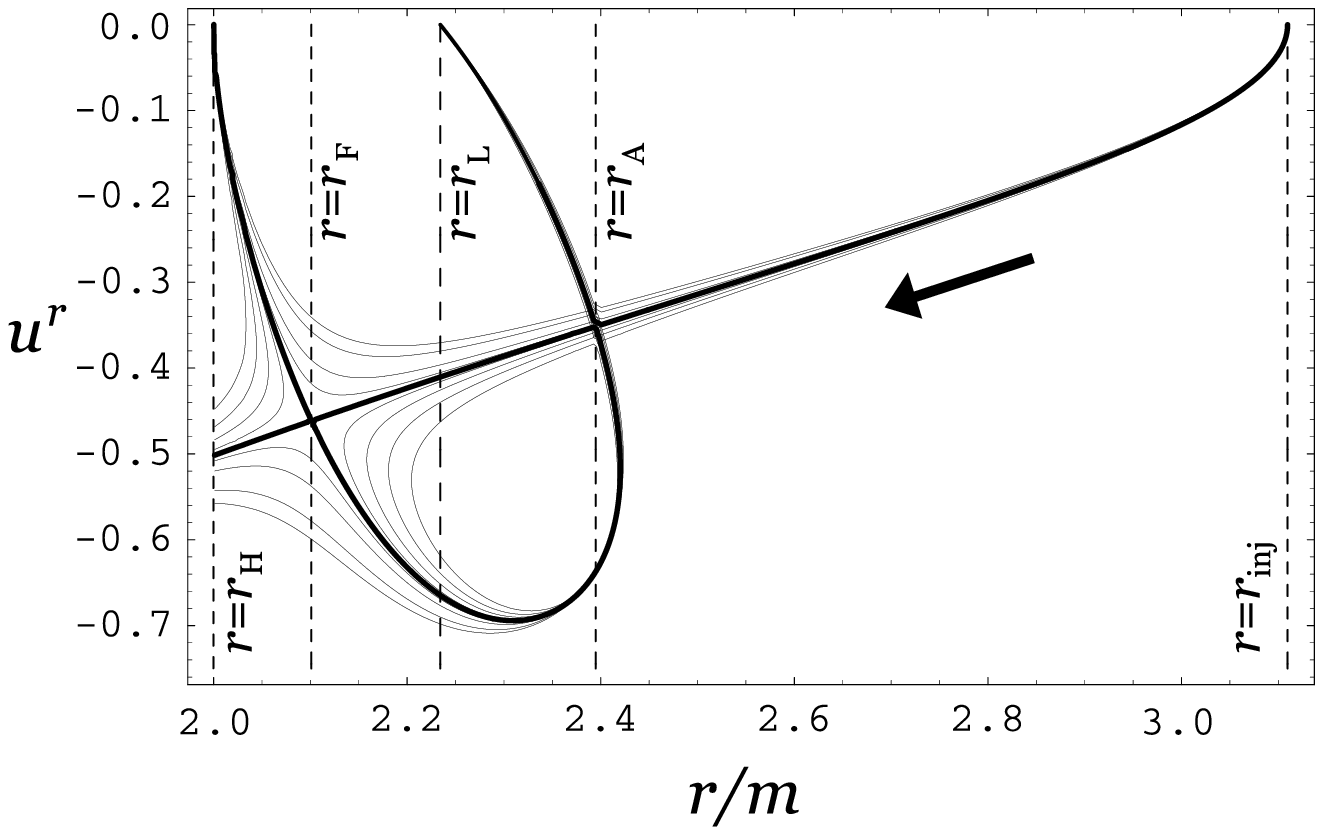}
    \caption{
 A preshock cold trans-fast MHD accretion solution
 ($\eta = \eta_{\rm F}$; bold curve with arrow). The solution with 
 zero-velocity at the event horizon $r=r_{\rm H}$ is unphysical.
 The other solutions ($\eta \neq \eta_{\rm F}$; thin curves) are also
 unphysical.
 The radii of $r=r_{\rm F}$, $r=r_{\rm L}$, $r=r_{\rm A}$ and
 $r=r_{\rm inj}$ are the locations of the fast magnetosonic point,
 the light surface, the Alfv\'en point and the injection point,
 respectively.
 A slow magnetosonic shock would be possible somewhere between 
 the plasma injection radius and the Alfv\'en radius.
 }
\label{fig:accretion}
\end{figure}

\placefigure{fig:accretion}

Figure~\ref{fig:accretion} illustrates solutions for the poloidal flow 
equation (\ref{eq:pol-eq}) for a radial field geometry in the equatorial 
plane, where the plasma is cold ($P=0$) with given parameters $E$, $L$ 
and $\Omega_F$ in the Schwarzschild spacetime. 
[Note that to determine the magnetic field configuration we should 
treat the equation of cross-field momentum balance (Grad-Shafranov 
equation); the general relativistic extension was obtained by 
\cite{ni91}. However, the task is very complicated, so that here 
we only use a given magnetic field configuration 
(see section 5).] 
The flow must pass through the fast magnetosonic point (TNTT90). 
That is possible only if there is some specific relation 
$\eta =\eta(E, L, \Omega_F)$; 
Only for the heavy solid curve with the arrow is this relation 
satisfied. The physical accretion solution starts from $r=r_{\rm inj}$ 
with zero poloidal velocity and reaches the event horizon 
$r=r_{\rm H}$ with a non-zero poloidal velocity.  
A slow magnetosonic shock should be located
between the plasma injection radius $r=r_{\rm inj}$ 
and the Alfv\'en radius $r=r_{\rm A}$, if the shock conditions are 
satisfied. After the slow magnetosonic shock, the heated postshock flow 
falls into the black hole after passing the slow magnetosonic point, 
the Alfv\'en point and the fast magnetosonic point, in this order. 
The radius of the light surface $r=r_{\rm L}$ and the Alfv\'en radius 
$r=r_{\rm A}$ do not change after the shock formation (see the next 
section).

\section{MHD Shocks in Kerr geometry}

\begin{figure}
    \epsscale{0.6}
    \plotone{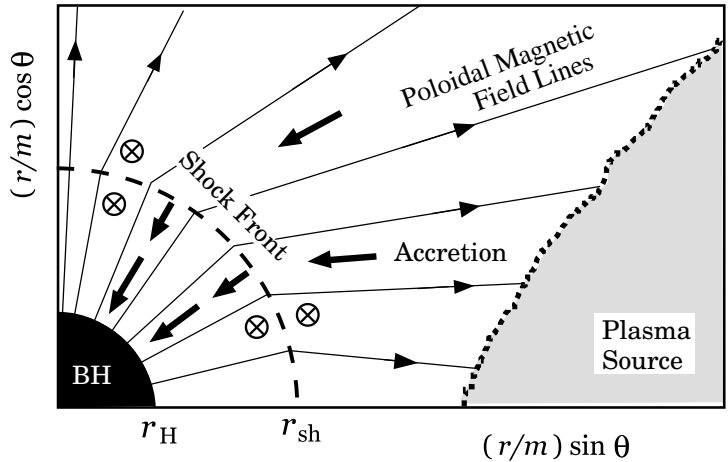}
    \caption{
 Accretion flows with shock front. It is assumed that in the poloidal 
 plane the downstream flow is radial and the shock front distributes 
 perpendicular to the the downstream flow direction. 
 In a general case, the magnetic field has a non-zero toroidal
 component (marked by $\otimes$) because of the plasma rotation. 
 }
\label{fig:sk-front}
\end{figure}

In order to explore the properties of MHD shocks associated with 
accretion flows near a black hole, we derive in this section the 
general relativistic version of MHD shock conditions --- an extension 
of the work by AC88.  Figure~\ref{fig:sk-front} shows a 
schematic picture of general accretion inflow from a plasma 
source, through a shock front, and onto a black hole. In this 
picture, the accretion originates from the plasma source (which 
can be the surface of a torus) rotating around a black hole. 
Strong shocks may be produced somewhere between the plasma source 
and the event horizon.

\placefigure{fig:sk-front}

\subsection{The Jump Conditions}

In a complete solution of the accretion that includes a shock, the flow 
must satisfy a set of conditions on either side of the discontinuity. 
The jump conditions for arbitrary shocks in a relativistic MHD flow are 
(e.g., AC88): 
\begin{eqnarray}
  &  [nu^\alpha] n_\alpha = 0 \ , & 
  \mbox{ --- the particle number conservation } \label{eq:nc} \\
  &  [T^{\alpha\beta}] n_\alpha = 0 \ , & 
  \mbox{ --- the energy momentum conservation } \label{eq:ec} \\
  &  [\mbox{\boldmath E}] \times \mbox{\boldmath n} = 0 \ , \ &
  \mbox{ --- the continuity relations for the electric field } \label{eq:En} \\
  &  [\mbox{\boldmath B}] \cdot \mbox{\boldmath n} = 0 \ , \ \  & 
  \mbox{ --- the continuity relations for the magnetic field } \label{eq:Bn}
\end{eqnarray}
where $n^\alpha=(n_0, \mbox{\boldmath n})$ and the square brackets 
denote the difference between the values of a quantity on the two 
sides of the shock.

We assume that the downstream flow velocity is radial (normal to the 
event horizon) and the shock front is perpendicular to the downstream 
flow {\boldmath n}=(1,0,0) as shown in Figure~2. Then, we set 
\begin{eqnarray}
  u^\alpha_1 &=& (u^t_1, u^r_1, u^\theta_1, u^\phi_1) \ , \\
  u^\alpha_2 &=& (u^t_2, u^r_2, 0 , u^\phi_2) \ , \\
  B^\alpha_1 &=& (B^r, B^\theta_1, B^\phi_1) \ , \\
  B^\alpha_2 &=& (B^r, 0 , B^\phi_2) \ , \\
  E^\alpha_1 &=& (E^r_1, E^\theta, 0 ) \ , \\
  E^\alpha_2 &=& (0 , E^\theta, 0 ) \ ,
\end{eqnarray}
where equations~(\ref{eq:En}) and (\ref{eq:Bn}) have been used.  
The subscripts ``1'' and ``2'' denote the preshock and the postshock  
quantities, respectively. From equation~(\ref{eq:Et}), we see that  
$\Omega_F$ does not change across the shock. Equations (\ref{eq:nc}) 
and (\ref{eq:ec}) evaluated in the shock rest frame yield the following 
relations 
\begin{equation}
  n_1 u^r_1 = n_2 u^r_2 \ , \label{eq:nc-sk}
\end{equation}
\begin{eqnarray}
  &\ & n_1 \mu_1 (u^r u_r)_1 -P_1 + \frac{1}{8\pi g_{tt}}
  ( -E^r_1 E_{r1} +B^\theta_1 B_{\theta 1} +B^\phi_1 B_{\phi 1})
  \nonumber \\
  &=& n_2 \mu_2 (u^r u_r)_2 -P_2 + \frac{1}{8\pi g_{tt}}
  ( B^\phi_2 B_{\phi 2}) \ , \label{eq:T=rr}
\end{eqnarray}
\begin{eqnarray}
  n_1 \mu_1 u^r_1 u^\theta_1 - \frac{1}{4\pi g_{tt}}
  \left( B^r B^\theta_1 + E^r_1 E^\theta \right) &=& 0 \ , 
                                                      \label{eq:T=rtheta}\\
  n_1 \mu_1 u^r_1 u^\phi_1 - \frac{B^\phi_1 B^r}{4\pi g_{tt}}
  &=& n_2 \mu_2 u^r_2 u^\phi_2 - \frac{B^\phi_2 B^r}{4\pi g_{tt}}\ ,  
                                                      \label{eq:T=rphi}\\
  n_1 \mu_1 u^r_1 u^t_1 - \frac{1}{4\pi g_{tt}}
  \left( \frac{B_{\phi 1} E_\theta}{\sqrt{-g}} + B^t_1 B^r \right)
  &=& n_2 \mu_2 u^r_2 u^t_2 - \frac{1}{4\pi g_{tt}}
  \left( \frac{B_{\phi 2} E_\theta}{\sqrt{-g}} + B^t_2 B^r \right) \ .
                                                         \label{eq:T=rt}
\end{eqnarray}
From the MHD conditions, we also obtain 
\begin{eqnarray}
  & & E_{r1} u_{t1} + \sqrt{-g}(B^\phi_1 u^\theta_1 - B^\theta_1 u^\phi_1)
  =0 \ , \label{eq:MHD-1}\\
  & & (B^r u^\phi_1 - B^\phi_1 u^r_1)\frac{1}{u_{t1}}
  =\frac{-E_\theta}{\sqrt{-g}}
  =(B^r u^\phi_2 - B^\phi_2 u^r_2)\frac{1}{u_{t2}} \ ,    \label{eq:MHD-2}\\
  & & (g_{t\phi}E_\theta + \sqrt{-g}B^r)u^\theta_1
  +(g_{t\phi}E_{r1} -\sqrt{-g}B^\theta_1) u^r_1 =0 \ ,    \label{eq:MHD-3}\\
  & & E_{r1} u^r_1 + E_{\theta 1} u^\theta_1 = 0 \ .      \label{eq:MHD-0}
\end{eqnarray}

From the shock condition of $(T^{rt})_1=(T^{rt})_2$, we obtain 
\begin{equation}
  (E-\omega L)_1 = (E-\omega L)_2 \ \ \ ,
\end{equation}
where $\omega \equiv -g_{t\phi}/g_{\phi\phi}$ is the angular velocity 
of the zero angular momentum observer (ZAMO) with respect to a distant 
observer, and from the shock condition of $(T^{r\phi})_1=(T^{r\phi})_2$, 
we also obtain 
\begin{equation}
  (g_{t\phi}E+g_{tt}L)_1 = (g_{t\phi}E+g_{tt}L)_2 \ \ \ .
\end{equation}
From these relations, we see that $E_1=E_2$ and $L_1=L_2$.  Furthermore, 
from the particle number conservation and the continuity for electric 
and magnetic fields at the shock front, we also obtain the relations 
$(\Omega_F)_1=(\Omega_F)_2$ and $\eta_1=\eta_2$. Thus, the field-aligned 
flow parameters $E$, $L$, $\Omega_F$ and $\eta$ are conserved across the 
shock front. This means that the location of the light surface and the 
Alfv\'en radius do not change by a shock generation.  The locations of 
light surfaces $r=r_{\rm L}$ are given by $\alpha=0$ and the Alfv\'en 
radius $r=r_{\rm A}$ is given by $\tilde L=-(G_\phi/G_t)_{\rm A}$. 
Because of the entropy generation ($S_2>S_1$), the fast/slow 
magnetosonic points appear at different locations in the preshock and 
postshock flow solutions. The increasing entropy makes the Alfv\'en wave 
speed at the Alfv\'en point $u_{\rm A}$ [$\equiv u_{\rm AW}(r_{\rm A})$] 
in a postshock solution smaller than that in a preshock solution.

\subsection{Dimensionless Parameters and Their Relations}

From the energy momentum conservation at the shock front, we have 
$(T^{rt})_1=(T^{rt})_2$, where $T^{rt}$ can be reduced to 
\begin{eqnarray}
  T^{rt} &=& g^{tt} ({\cal E}^r_{\rm fluid}+{\cal E}^r_{\rm em})
      -g^{t\phi}({\cal L}^r_{\rm fluid}+{\cal L}^r_{\rm em})
                                                      \label{eq:Trt-a} \\
  &=&  g^{tt} \left[ n \mu u^r u_t (1-\omega\ell)
     + \frac{B_\phi B^r(\Omega_F-\omega)}{4\pi G_t}\right] \ ,
                                                      \label{eq:Trt-b}
\end{eqnarray}
and $\ell\equiv -u_\phi/u_t$ is the specific angular momentum of the 
plasma. Here, we define dimensionless parameters. First, we define the 
``magnetization parameter'', which denotes the ratio of the Poynting 
flux to the total mass-energy flux seen by ZAMO, 
\begin{equation}
  \sigma \equiv \frac{({\cal E}^r -\omega{\cal L}^r)_{\rm em} }
        {({\cal E}^r -\omega{\cal L}^r)_{\rm fluid}}\ ,
\end{equation}
then we can express equation~(\ref{eq:Trt-a}) as 
\begin{equation}
  T^{rt} = g^{tt}({\cal E}^r -\omega{\cal L}^r)_{\rm fluid}(1+\sigma)
  = (nu^r)(\mu u^t)(1+\sigma) \ .                    \label{eq:Trt-c}
\end{equation}
Then, the magnetization parameter can be reduced to 
\begin{equation}
  \sigma = \frac{B_\phi}{4\pi\eta(\mu u^t)} \frac{G_\phi}{\rho_w^2} \ .
  \label{eq:sigma-1}
\end{equation}
By using (\ref{eq:ut}) and (\ref{eq:ufai}), $\sigma$ is denoted 
in terms of $M^2$, $E$, $L$ and $\Omega_F$, as 
\begin{equation}
  \sigma = - \frac{ (G_\phi E + G_t L) G_\phi }
  {\rho_w^2 e + (g_{\phi\phi}E + g_{t\phi}L)M^2}
  = - \frac{e-h\alpha}{e-hM^2} \ , \label{eq:sigma-2}
\end{equation}
where $h \equiv g^{tt}(E-\omega L)$. We can also express $M^2$ in 
terms of $\sigma$, $E$, $L$ and $\Omega_F$.

Next, we define the following dimensionless parameters 
\begin{eqnarray}
  q & \equiv & \frac{B^{\phi}_{2}}{B^{\phi}_{1}} \
     = \ \frac{M^2_1-\alpha}{M^2_2-\alpha} \ ,          \label{eq:q} \\
  \zeta & \equiv & \frac{\mu_2 u^t_2}{\mu_1 u^t_1} \
     = \ \frac{e-hM^2_2}{e-hM^2_1}q \ ,                 \label{eq:zeta} \\
  \xi & \equiv & \frac{n_2u^t_2}{n_1u^t_1} \
     = \ \frac{M^2_1}{M^2_2}\zeta \ ,                   \label{eq:xi}
\end{eqnarray}
where $q$ is the amplification factor for the toroidal magnetic 
field and $\xi$ is the shock frame compression ratio.  From 
equations~(\ref{eq:nc-sk}), (\ref{eq:T=rt}), (\ref{eq:MHD-2}) 
we obtain the following relation 
\begin{equation}
  1-\zeta = \sigma_1 \left( q -1 \right) \ . \label{eq:ze-si}
\end{equation}

The normalization of the 4-velocity gives an equation for $u^t_2$ 
\begin{equation}
  u^t_2=\left[ g_{tt} +2g_{t\phi}\Omega_2 +g_{\phi\phi}\Omega_2^2
       +\frac{(u^ru_r)_1}{\xi^2(u^t_1)^2} \right]^{-1/2}
        \ , \label{eq:ut2}
\end{equation}
where $\Omega\equiv u^\phi/u^t$ is the angular velocity of the fluid, 
and 
\begin{equation}
  \Omega_2 =
      \frac{e\Omega_F-\rho_w^{-2}(g_{t\phi}E+g_{tt}L) M^2_2}{e-hM^2_2} \ .
      \label{eq:om2b}
\end{equation}
When other postshock quantities $\xi$ or $M^2_2$ are determined, $u^t_2$ 
and $\Omega_2$ can be obtained.

From equation~(\ref{eq:T=rr}), we obtain 
\begin{equation}
  1-\frac{\zeta}{\xi}-\frac{P_1-P_2}{n_1\mu_1 (u^ru_r)_1}
  + \frac{(-E^rE_r+B^\theta B_\theta +B^\phi B_\phi)_1-(B^\phi B_\phi)_2}
  {8\pi g_{tt} n_1\mu_1 (u^ru_r)_1 } = 0 \ .
\end{equation}
Using the relations 
\begin{eqnarray}
  -E^rE_r+B^\theta B_\theta &=&
  \frac{g_{tt}\alpha}{G_t^2} B^\theta B_\theta \ , \\
  (B^\phi B_\phi)_1- (B^\phi B_\phi)_2 &=&
  (B^\phi B_\phi)_1 (1-q^2) \ ,
\end{eqnarray}
we get 
\begin{equation}
  1 - \frac{\zeta}{\xi} = \Pi - {\cal X}_1 + (q^2-1){\cal T}_1
\end{equation}
with 
\begin{eqnarray}
  \Pi &\equiv& \frac{P_1-P_2}{n_1\mu_1(u^ru_r)_1}
  \ =\ 1-\frac{1}{\xi}+\frac{\sigma_1}{\xi}(q-1)
  + {\cal X}_1 - (q^2-1){\cal T}_1 \ , \label{eq:pi} \\
  {\cal X}_1 &\equiv& \frac{\alpha (B^\theta B_\theta)_1}
  {2 B^rB_r M^2_1} \ , \\
  {\cal T}_1 &\equiv& \frac{\sigma_1}{2}\left(\frac{u_t}{u^r}\right)_1^2
  \frac{\rho_w^2(\Omega_1-\Omega_F)}{g_{rr}G_\phi} \ ,
\end{eqnarray}
which depend on $\xi$, $q$ and upstream parameters $M^2_1$ ( or 
$\sigma_1$ ), $E$, $L$ and $\Omega_F$. The relation between $M^2_2$ 
and $M^2_1$ are given by the poloidal equation (\ref{eq:pol-eq}).  
(At the shock location $r=r_{\rm sh}$, both $M^2=M^2_1$ and $M^2=M^2_2$ 
are solutions of the poloidal equation.)

In the following, for preshock accretion flows we restrict ourselves 
to the cold limit ($P_1=0$).  Using the definition of $\zeta$ and the 
equation of state for a Boltzmann gas with the polytropic index $\Gamma$ 
\begin{eqnarray}
  \mu_1 &=& m_{\rm p} \\
  \mu_2 &=& m_{\rm p} \left( 1+\frac{\Gamma}{\Gamma-1}
  \frac{P_2}{n_2 m_{\rm p}} \right) \ ,
\end{eqnarray}
we find with equation~(\ref{eq:pi}) 
\begin{equation}
  1 -\frac{u^t_2}{u^t_1}=\sigma_1(q-1) - \frac{\Gamma}{\Gamma-1}
  \frac{g_{rr}(u^r/u^t)_1^2(u^t_2)^2}{\xi} \Pi \ . \label{eq:xi0}
\end{equation}
Combining equations~(\ref{eq:q}), (\ref{eq:xi}) and (\ref{eq:ze-si}) 
gives the quadratic equation for $\zeta$ 
\begin{equation}
  \zeta^2 - \left[ 1 + \sigma_1 +\left( \frac{G_t}{M^2_1}
  - \frac{\sigma_1\Omega_F}{\Omega_F-\Omega_1} \right) \xi \right] \zeta
  + \left( \frac{G_t}{M^2_1}
  - \frac{\sigma_1\Omega_1}{\Omega_F-\Omega_1} \right) \xi = 0
                                             \ , \label{eq:zeta-quad}
\end{equation}
where we use the relation $\Omega-\Omega_F=\sigma M^2/G_\phi$. 
We are now able to eliminate $u^t_2$ and $q$ from equation~(\ref{eq:xi0}). 
After considerable manipulations we obtained a polynomial of eighth 
degree in $\xi$ 
\begin{equation}
  \sum_{i=0}^8 c_i(M^2_1, \Gamma ; m, a, E, L, \Omega_F, \eta)\xi^i = 0 \ .
  \label{eq:ci7}
\end{equation}
The coefficients $c_i$ are dependent only on upstream parameters, 
except for $\Gamma$ which is a function of the downstream temperature 
involving the modified Bessel functions~\citep{sy57} 
\begin{equation}
  \Gamma(\Theta) = 1+ \left[\frac{1}{\Theta}\left(
  \frac{K_1(1/\Theta)}{K_2(1/\Theta)}-1
  \right) +3 \right]^{-1} \label{eq:gamma}
\end{equation}
with $\Theta=kT/m_{\rm p}$. $\Theta$ is related to $\xi$ though 
\begin{equation}
  \Theta = -g_{rr}u^t_1 u^t_2 (u^r_1/u^t_1)^2 \Pi/\xi \ .
\end{equation}
Thus, the compression ratio $\xi$ is the solution of a polynomial 
(\ref{eq:ci7}), which has to be solved simultaneously with equation 
(\ref{eq:gamma}) for the polytropic index $\Gamma$ for the shocked plasma. 
The concrete expressions of the coefficients $c_i$ are quite lengthly, 
so that we only show numerical results in the next section. 
This polynomial of eighth degree has in general several real solutions 
corresponding to the different shock transitions. The downstream 
quantities $\zeta$, $q$, $u^t_2$, $u^r_2$, $\Omega_2$ and $\Pi$ are 
also obtained from equations (\ref{eq:zeta-quad}), (\ref{eq:q}), 
(\ref{eq:ut2}), (\ref{eq:nc-sk}), (\ref{eq:om2b}) and (\ref{eq:pi}), 
respectively.

\section{Results and Discussions for Slow Magnetosonic Shocks }

To see the general shock behavior which depends on flow parameters with 
the hole's spin $a$, we present solutions to the coupled equations 
(\ref{eq:ci7}) and (\ref{eq:gamma}) for a range of black hole spins. 
For computational reasons, instead of equation~(\ref{eq:gamma}) we use 
a polynomial approximation given by \cite{se86}.  
In this paper, we restrict ourselves to slow magnetosonic shocks, which 
should be located between the plasma source and the Alfv\'en point. 
Then, we will only treat the sub-Alfv\'enic region of $M < M_{\rm A}$  
on a trans-fast MHD accretion solution, which is considered as a 
preshock accretion solution and is assumed cold. This preshock 
accretion is super-slow magnetosonic. 
The postshock accretion would be heated up at the shock front, and 
the slow magnetosonic wave in the hot plasma has a non-zero wave 
speed $(u_{\rm SM})_2 > 0$, where $u_{\rm SM}$ is the slow 
magnetosonic wave speed (see, TNTT90). Then, the hot postshock accretion 
must satisfy the condition $0< u_{p2} < (u_{\rm SM})_2$ at the shock 
location; that is, the postshock flow must be sub-slow magnetosonic. 
All our preshock accretion solutions presented in this paper refer to 
physically acceptable accretion flows, in the sense that the critical 
condition (as shown as an example in Fig.~\ref{fig:accretion}) is 
satisfied.

Since we have four conserved parameters, $E$, $L$, $\Omega_F$ and
$\eta$, two of these quantities are kept the same for all spin
parameters. (The physical descriptions of the parameters are summarized
in Table~\ref{tab:add-1}.)
Fixing $L$ and $\Omega_F$ would prevent the formation of shocks across 
the entire spin range, and so $E$ and $\eta$ were held constant. 
Then, physical parameter sets for the trans-fast MHD accretion 
solutions are calculated. 
The trans-fast MHD accretion solutions were then obtained by letting 
$\Omega_F$ increase and $L$ (and thus $e$) decrease linearly with 
increasing spin $a$ (Table~\ref{tab:para-1}). 
For simplicity, we set the flow to be in the equatorial plane 
of the radial field geometry     
$B^{r} = B_{\rm inj}G_t/\Sigma$ and $B^{\theta} = 0$ 
(and then $u^{\theta}=0$). 
$B_{\rm inj}$ is the strength of the field at the plasma injection
point. Here, we should note that the radii of 
$r = r_{\rm H} \equiv m+\sqrt{m^2-a^2}$, 
$r=r_{\rm F}$, $r=r_{\rm L}$, $r=r_{\rm A}$ and $r=r_{\rm inj}$ are 
different for each flow solutions. For larger spin $a$, the region of 
$r_{\rm A} < r < r_{\rm inj}$, where the slow magnetosonic shock is 
expected, becomes narrow and shifts inward (toward the horizon).

\placetable{tab:add-1}

\placetable{tab:para-1}

\begin{figure}
    \epsscale{0.45}
     \plotone{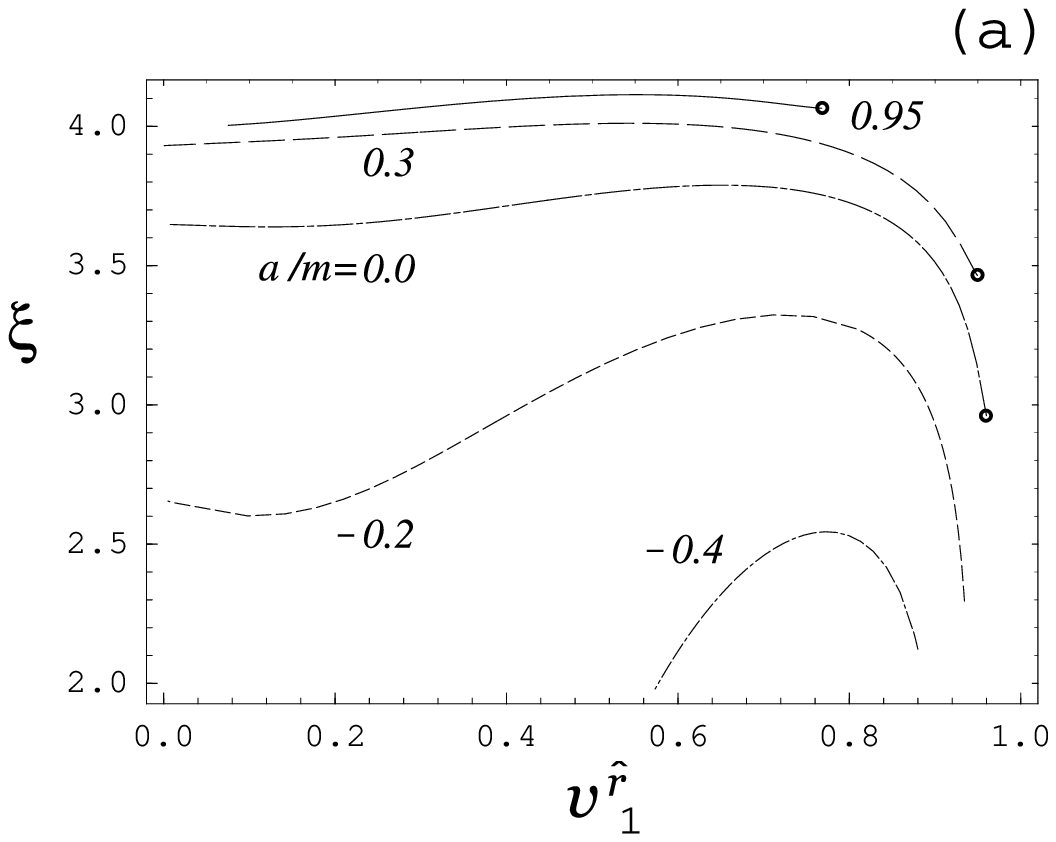}
     \plotone{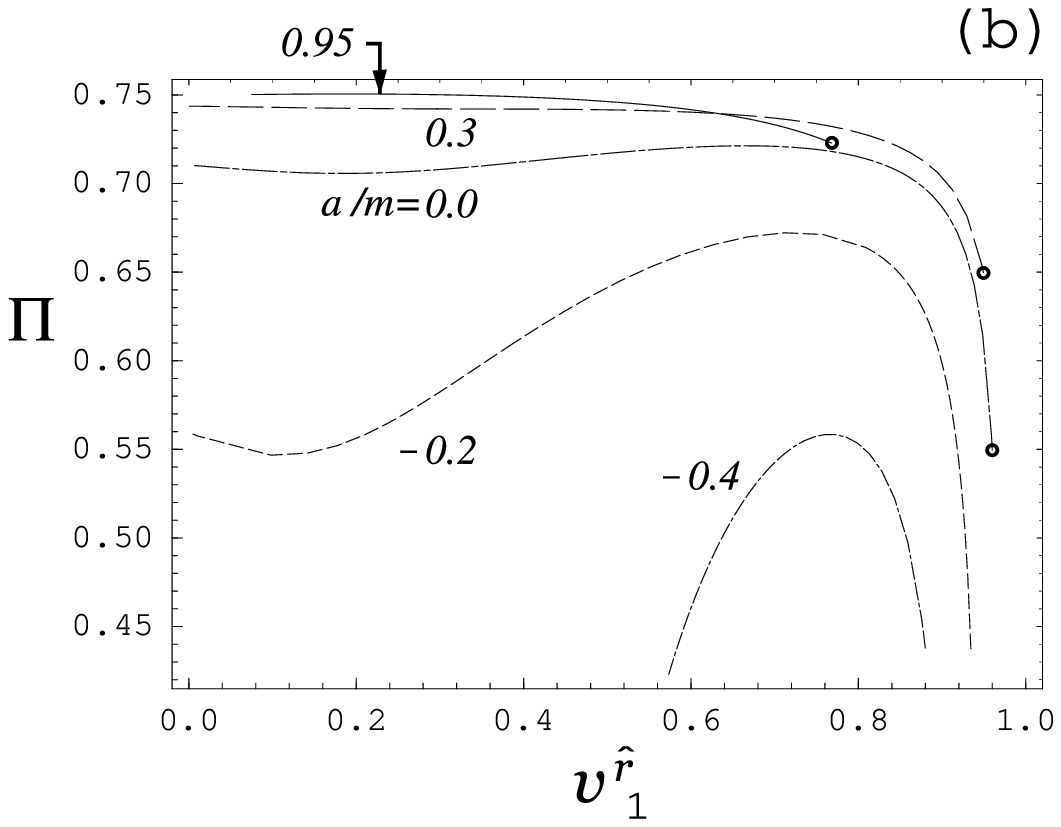}
     \plotone{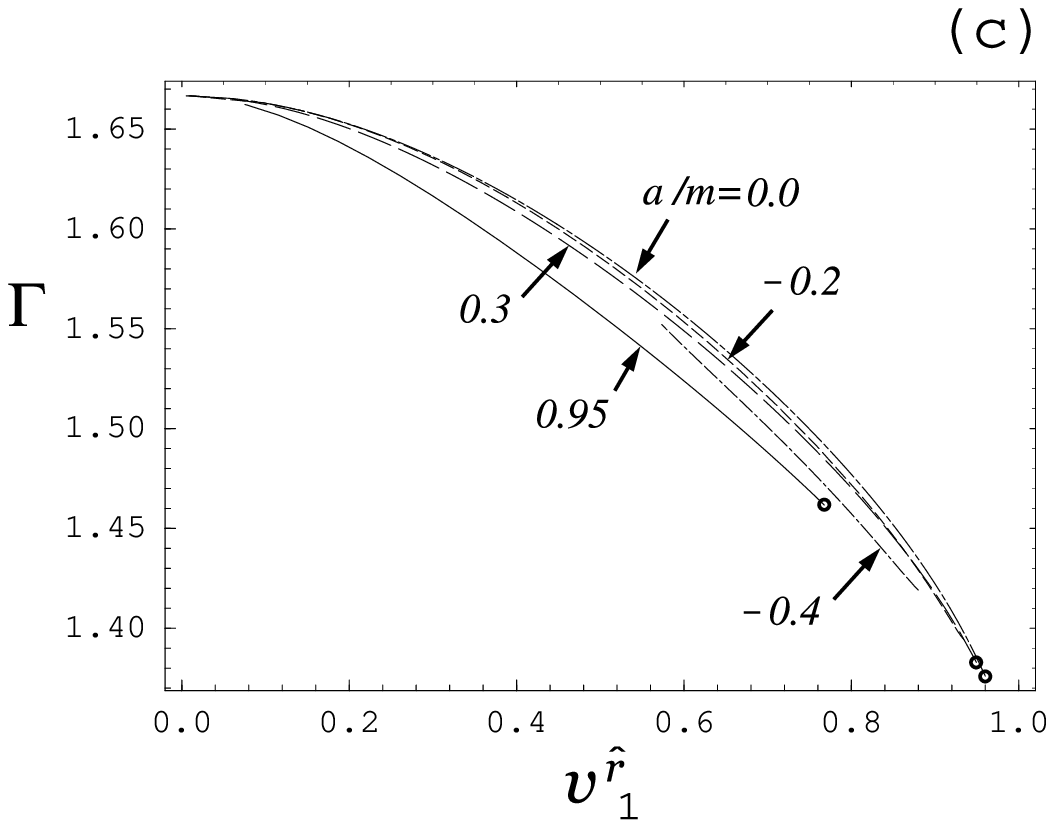}
     \plotone{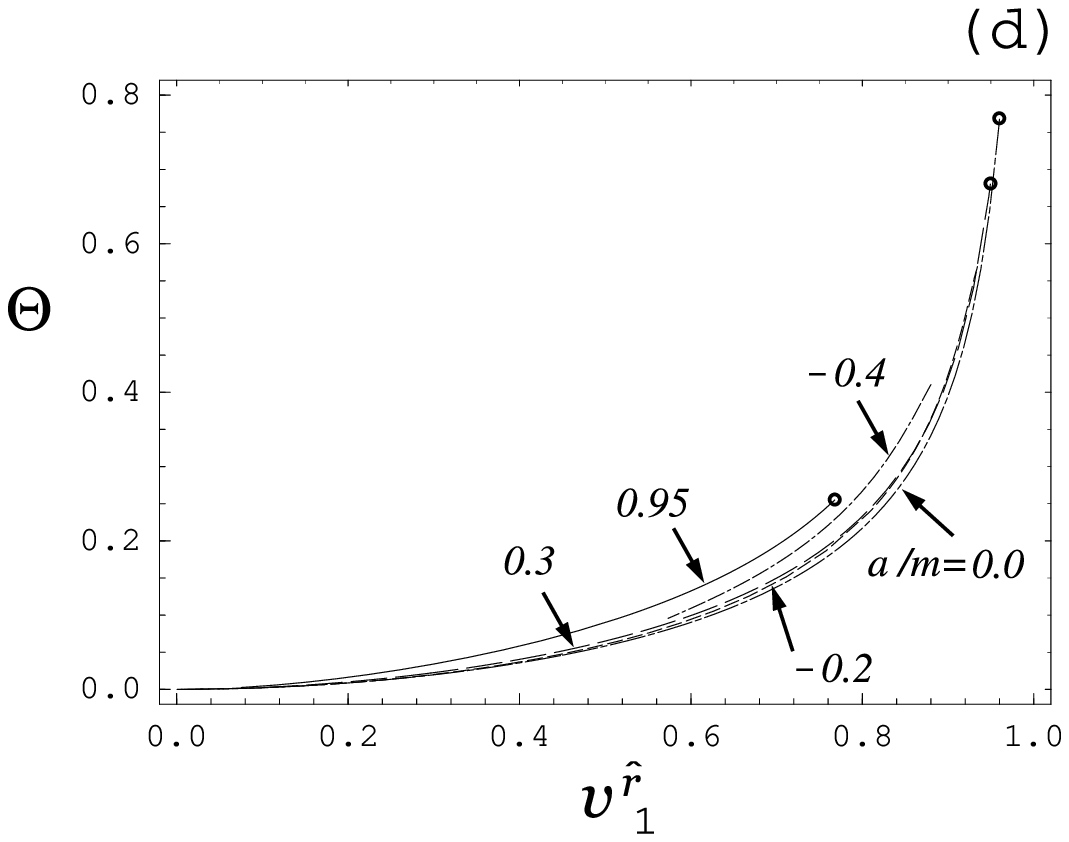}
    \caption{
 (a) The shock frame compression ratio $\xi$,
 (b) thermal pressure $\Pi$,
 (c) polytropic index $\Gamma$, and
 (d) temperature parameter $\Theta$, respectively,
 as a function of $v^{\hat r}_1$, the radial velocity of the infalling 
 plasma, for variable spin parameters ($a/m=0.95, 0.3, 0.0, -0.2, -0.4$).  
 The parameters $\Pi$, $\Gamma$ and $\Theta$ refer to the downstream
 values. The shocks at the switch-off points are shown by small spots. 
 The sets of field aligned parameters are given in
 Table~\ref{tab:para-1}. 
 }
\label{fig:sk-1}
\end{figure}

\placefigure{fig:sk-1}

\begin{figure}
    \epsscale{0.5}
    \plotone{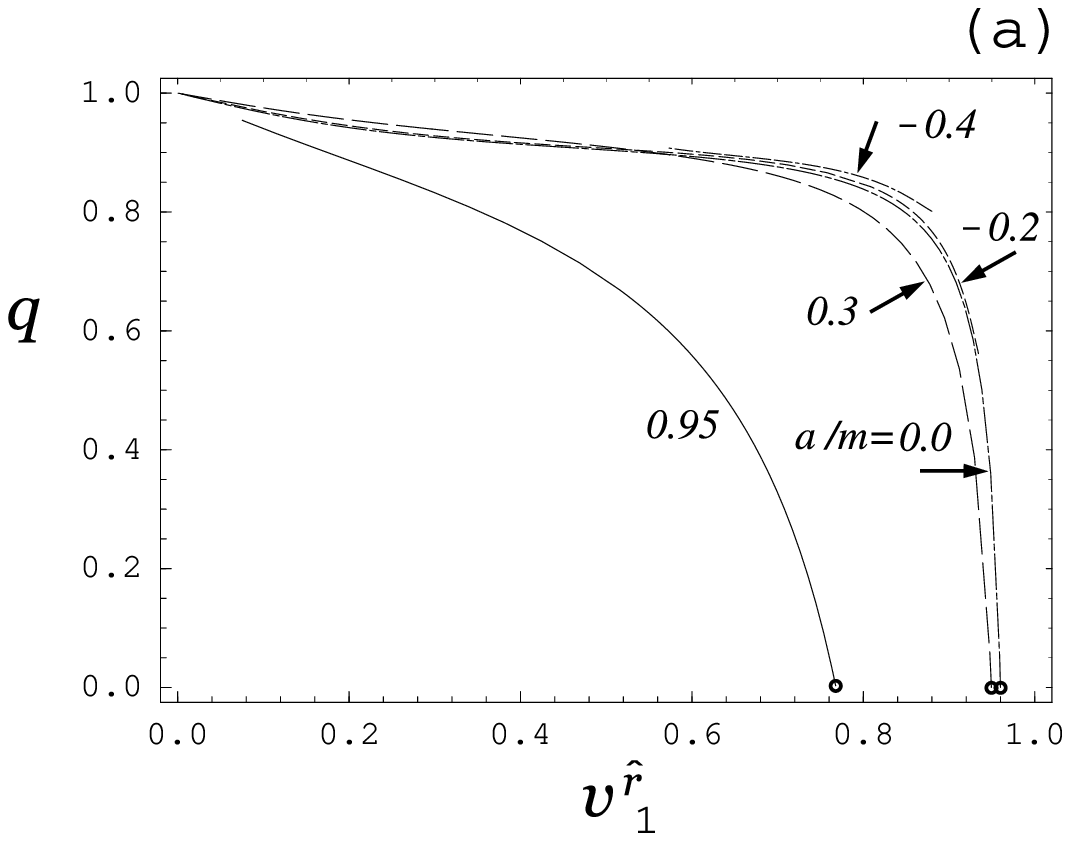}
    \plotone{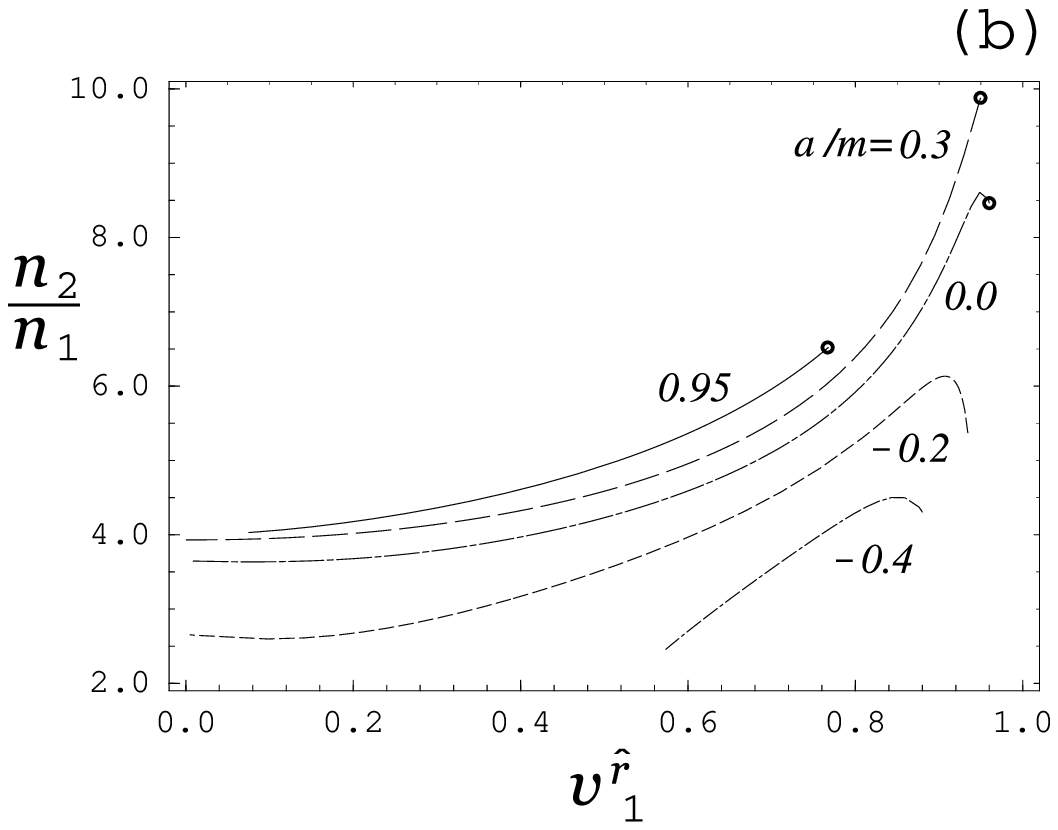}
    \plotone{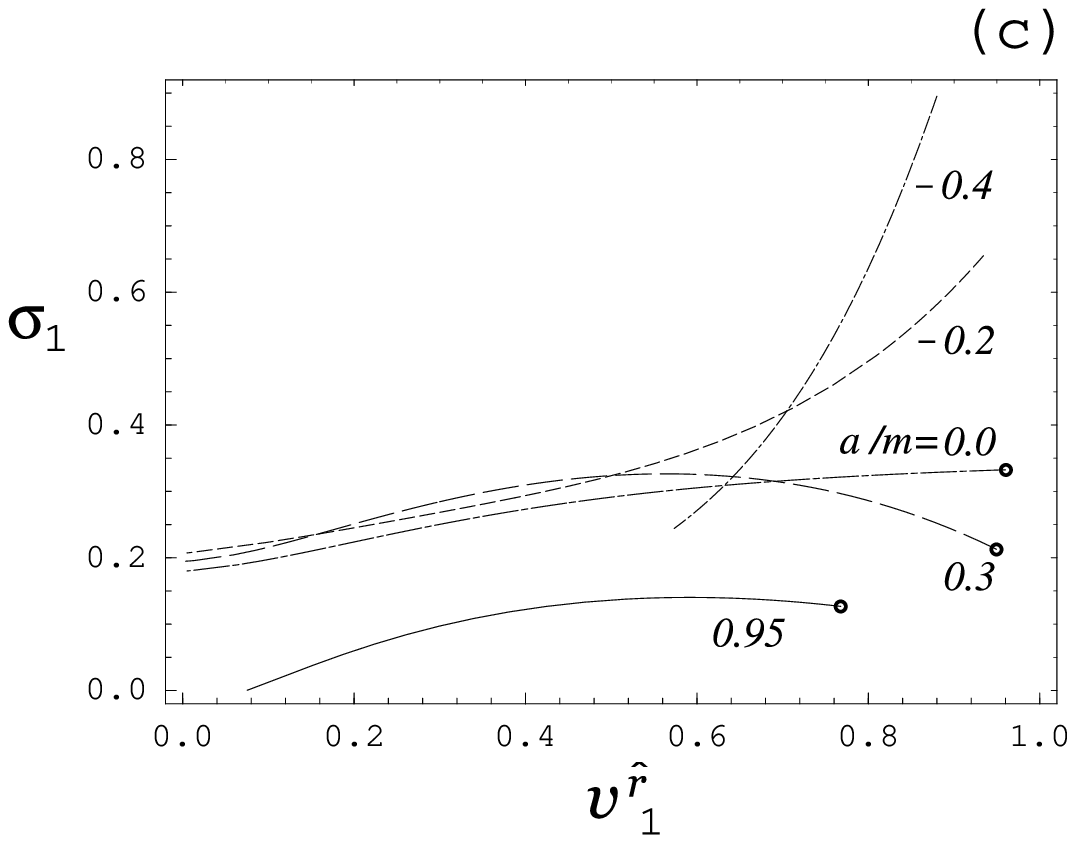}
    \caption{ 
  Various parameters are shown as a function of $v^{\hat r}_1$. 
 (a) The ratio of the downstream toroidal magnetic field to upstream
  toroidal magnetic field $q$,
 (b) the ratio of the downstream number density to upstream number 
 density,  and
 (c) the upstream magnetization parameter $\sigma_1$. 
 The chosen parameter sets are the same as in Fig.~\ref{fig:sk-1}. 
 }
\label{fig:sk-2}
\end{figure}

\placefigure{fig:sk-2}

In figures~\ref{fig:sk-1} and~\ref{fig:sk-2} we show the effect of 
black hole rotation on the nature of shock and accretion flows. 
The spin parameters selected are $a/m$ = 0.95, 0.3, 0.0, $-0.2$ 
and $-0.4$.  (In this paper we only calculate the cases 
$\Omega_F>\omega_{\rm H}$.)   
In the figures the solid curves are for the maximum $a$ 
($= 0.95m$), and the spin decreases as it changes from the solid 
to the dot-dashed curves which are the minimum at $a = -0.4m$. 
Positive $a$ refers to black hole and magnetosphere corotation, while 
negative $a$ refers to the counter-rotation. No solutions were found 
for $a < -0.4m$.  This is due to the choice of conserved quantities.

Figures~\ref{fig:sk-1}a,~\ref{fig:sk-1}b,~\ref{fig:sk-1}c 
and~\ref{fig:sk-1}d show the shock frame compression ratio $\xi$, 
thermal pressure $\Pi$, polytropic index $\Gamma$, and temperature 
parameter $\Theta$, respectively, as a function of $v^{\hat r}_1$, 
where $v^{\hat r} \equiv -(\sqrt{A}/\Delta)u^r/u^t$ is the positive 
radial 3-velocity  of the infalling plasma as seen by a ZAMO. 
The parameters $\Pi$, $\Gamma$ and $\Theta$ refer to the downstream 
values. Figure~\ref{fig:sk-1}a shows in general that the compression 
ratio $\xi$ increases and then decreases with velocity.  
Figures~\ref{fig:sk-1}a and~\ref{fig:sk-1}b show that both $\xi$ and 
$\Pi$ increase with increasing black hole rotation. Thermal energy 
$kT$ (or $\Theta$) and polytropic index $\Gamma$ for postshock flow 
have a weak dependence on the black hole spin, but the general trend 
is that for a given velocity, $\Gamma$ slightly decreases while 
$\Theta$ slightly increases with increasing $|a|$.  
Figures~\ref{fig:sk-2}a,~\ref{fig:sk-2}b and~\ref{fig:sk-2}c show 
the effect of velocity on three other interesting parameters --- 
the ratio of downstream toroidal magnetic field to upstream 
toroidal magnetic field $q$, the ratio of downstream number density 
to upstream number density $n_2/n_1$, and the upstream magnetization 
parameter $\sigma_1$.  We find that $q$ generally decreases while the 
ratio of the postshock to preshock density increases with the velocity. 
The effect of larger spin is to enhance the density 
jump while the azimuthal magnetic field amplification and 
magnetization are suppressed. 
In Figure~\ref{fig:sk-2}a, we can see the switch-off shock, except for 
the  counter-rotating black hole cases. The switch-off shock occurs in 
the  limit when the flow velocity equals the Alfv\'en wave speed at 
$r_{\rm sh}=r_{\rm A}$. At this limit, for the postshock flow the 
toroidal magnetic field $B^\phi_2$ seen by distant observers becomes 
zero ($q=0$). When the shock happens close to the Alfv\'en point, 
a very hot plasma region is generated (the small spots in 
Fig.~\ref{fig:sk-1}d). This region could emit X-rays and $\gamma$-rays 
around the black hole.

In the Newtonian analogy, we expect that for the strong shock 
$n_2/n_1 \sim 4$ for $\Gamma=5/3$ and $n_2/n_1 \sim 7$ for 
$\Gamma=4/3$. In the Schwarzschild black hole case, for example, 
we obtain a similar tendency for the maximum values; that is, 
$n_2/n_1 \sim 4$ for $\Gamma=5/3$ and $n_2/n_1 \geq 8$ for 
$\Gamma=1.4$ (see Figs.~\ref{fig:sk-1}c and~\ref{fig:sk-2}b).  
The compression ratio also has a maximum $\xi_{\rm max} \approx 3.6$ 
for $a=0$ and $\xi_{\rm max}\geq 4$ for $a=0.95m$ (Fig.~\ref{fig:sk-1}a).  
Compared  with the $n_2/n_1$ vs $v^{\hat r}_1$ diagram, 
$\xi(v^{\hat r}_1)$ is almost constant for the strong shocks 
(i.e., $a>0$). This is due to the effect of factor $u^t_2/u^t_1$ 
(see, eq.~(\ref{eq:xi})).  
From Figures~\ref{fig:sk-1}a and~\ref{fig:sk-2}b,  
it may appear that a strong shock is possible for very low accretion 
speed (for $a > -0.2m$). Note that in our calculation, we set $\eta=$ 
constant ($\eta\propto nu^r \neq 0$), so $v^{\hat r}_1\ll 1$ means 
$n_1\gg 1$. When $v^{\hat r}_1\ll 1$, the postshock flow must also  
satisfy $v^{\hat r}_2\ll 1$, and then $n_2\gg 1$. Thus, as a result of 
numerical calculation, we obtain a finite value for $n_2/n_1$ in the 
limit of $v^{\hat r}_1\to 0$.

\begin{figure}
    \epsscale{0.5}
    \plotone{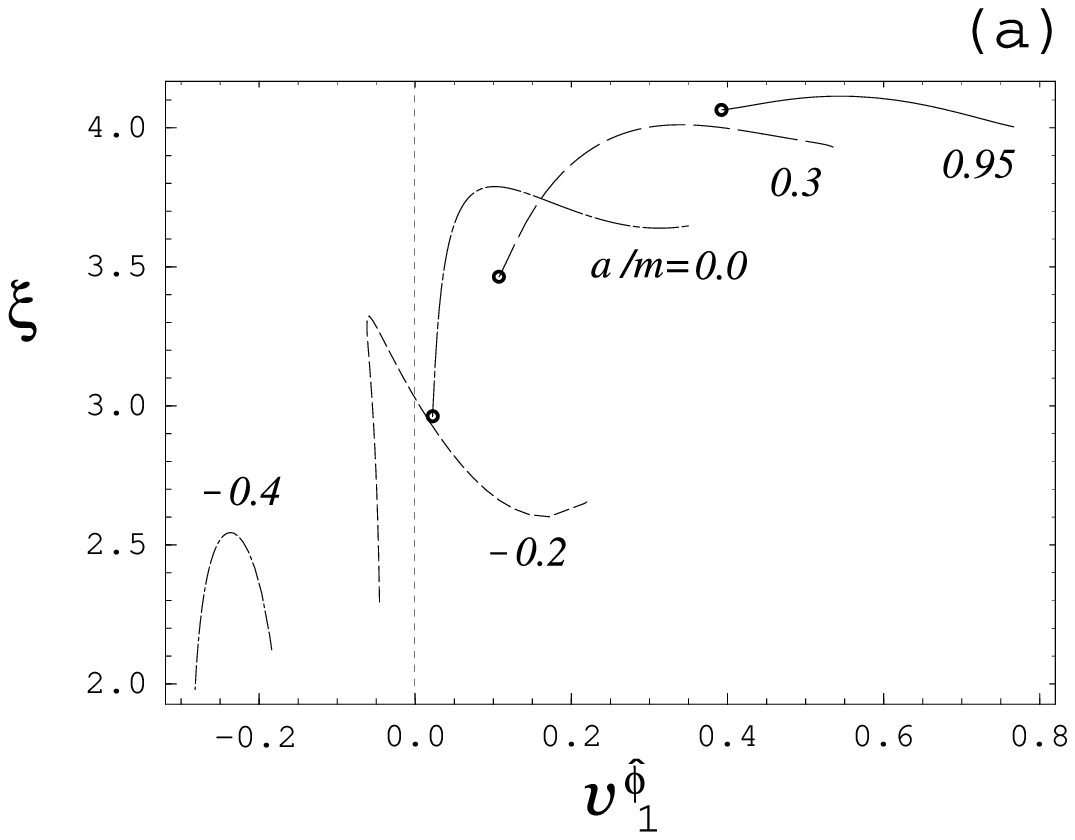}
    \plotone{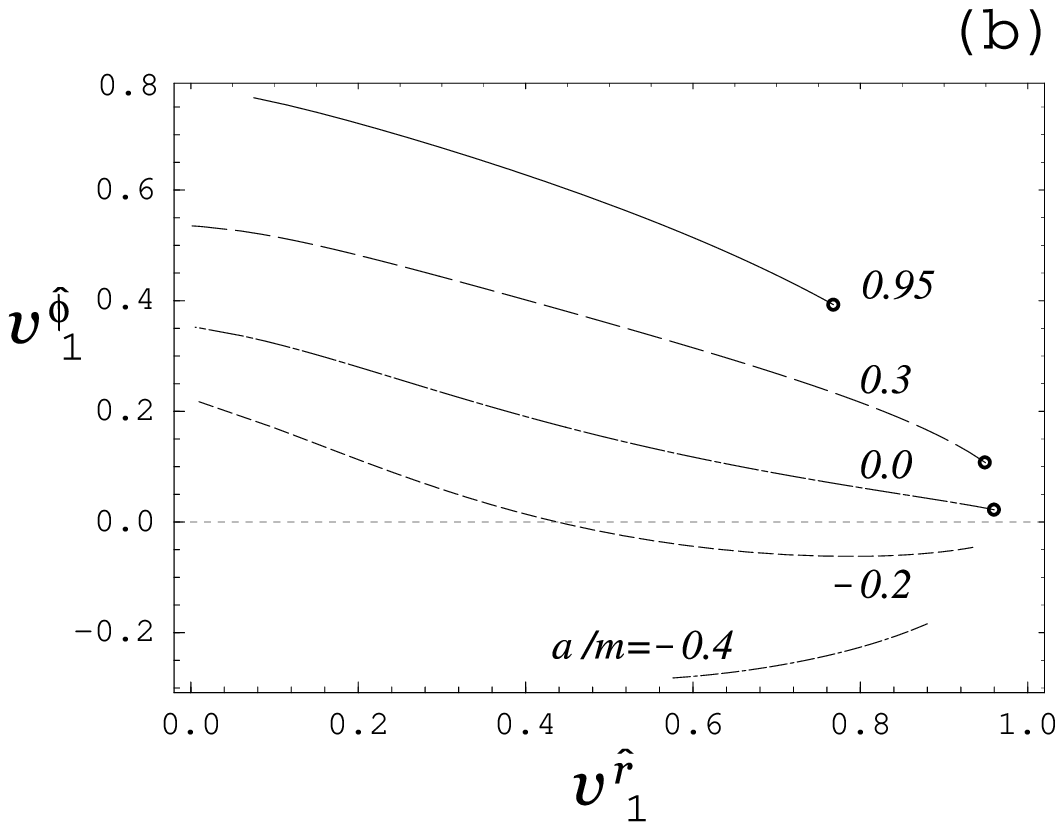}
    \plotone{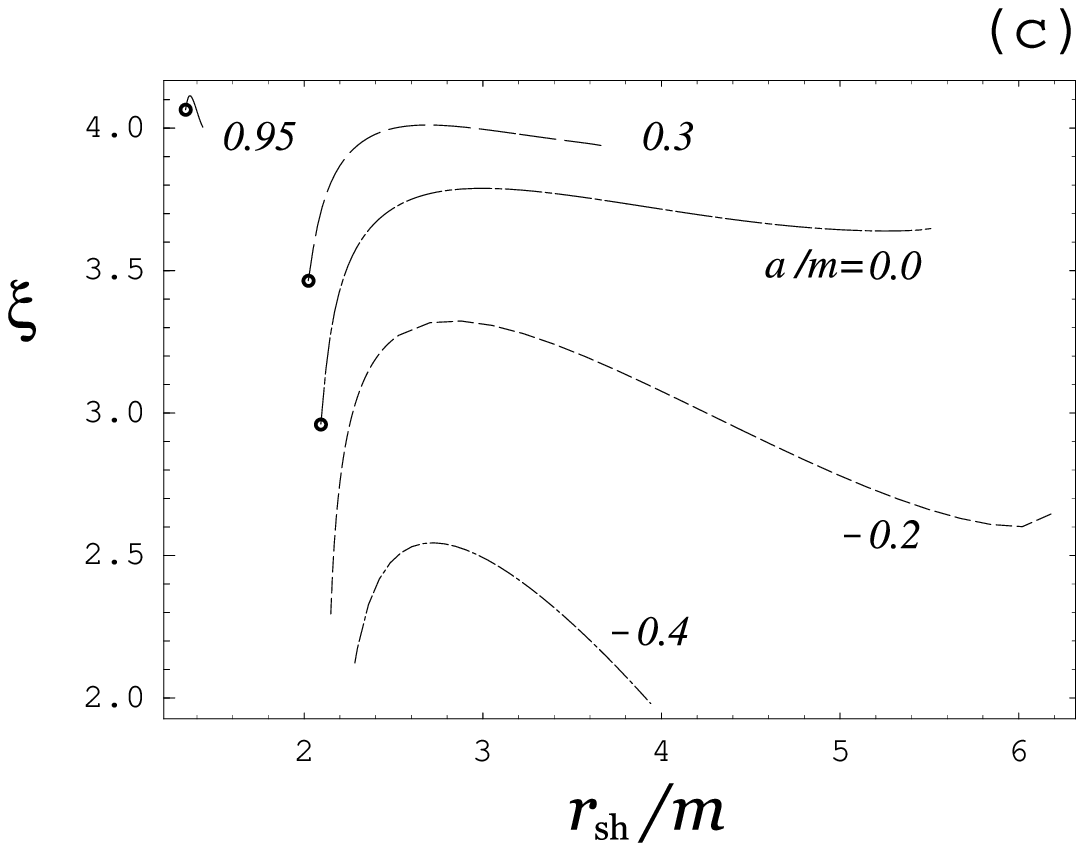}
    \caption{
 (a) The compression ratio as a function of $v^{\hat \phi}_1$,
 (b) the toroidal velocity  $v^{\hat \phi}_1$ as a function of 
 $v^{\hat r}_1$, and
 (c) the compression ratio $\xi$ as a function of the shock location
 $r_{\rm sh}$. The parameter sets are the same as in
 Fig.~\ref{fig:sk-1}. 
 }
\label{fig:sk-3}
\end{figure}

\placefigure{fig:sk-3}

Figures~\ref{fig:sk-3}a and~\ref{fig:sk-3}b show the $\xi$ vs. 
$v^{\hat \phi}_1$ and $v^{\hat \phi}_1$ vs. $v^{\hat r}_1$ relations, 
where $v^{\hat \phi} \equiv (-g_{\phi\phi}/\rho_w)(\Omega-\omega)$ 
is the toroidal 3-velocity seen by a ZAMO. Figure~\ref{fig:sk-3}b shows 
that when its radial velocity is very low, such a flow does possess 
sufficiently large toroidal velocity (especially the $a=0.95m$ case). 
The negative $v^{\hat \phi}_1$ is due to the effect of dragging on the 
plasma by the counter-rotating black hole. 
The toroidal velocity $v^{\hat\phi}_1$ is explicitly related to the 
toroidal magnetic field $(B_\phi)_1$, and therefore it can be also 
related to the magnetization parameter $\sigma_1$ 
(see Fig.~\ref{fig:sk-2}c). 
The difference of $\sigma_1$  between the corotating black hole case 
and the counter-rotating black hole case is caused by this
spin-dependence of the toroidal motion. That is, due to the dragging  
effects, $|B_\phi|_1$ is forced to decrease for $a=0.95m$, while it 
is forced to increase by stretching toward the ($-\phi$)-direction for 
the $a=-0.4m$ case. 
We also obtain the result that $(v^{\hat\phi}_2-v^{\hat\phi}_1) \to 0$ 
and the magnetic field line is tangent to the shock surface when 
$v^{\hat r}_1\to 0$. Figure~\ref{fig:sk-3}c shows the $\xi$ vs. 
$r_{\rm sh}$ relation.  For a magnetosphere with a large spin the shock 
front is located near the event horizon, and a strong shock is 
generated. Along each curve the shock strength increases slightly 
when the shock location shifts inward from the vicinity of the 
plasma source, but near the Alfv\'en point it decreases considerably.

\begin{figure}
    \epsscale{0.7}
    \plotone{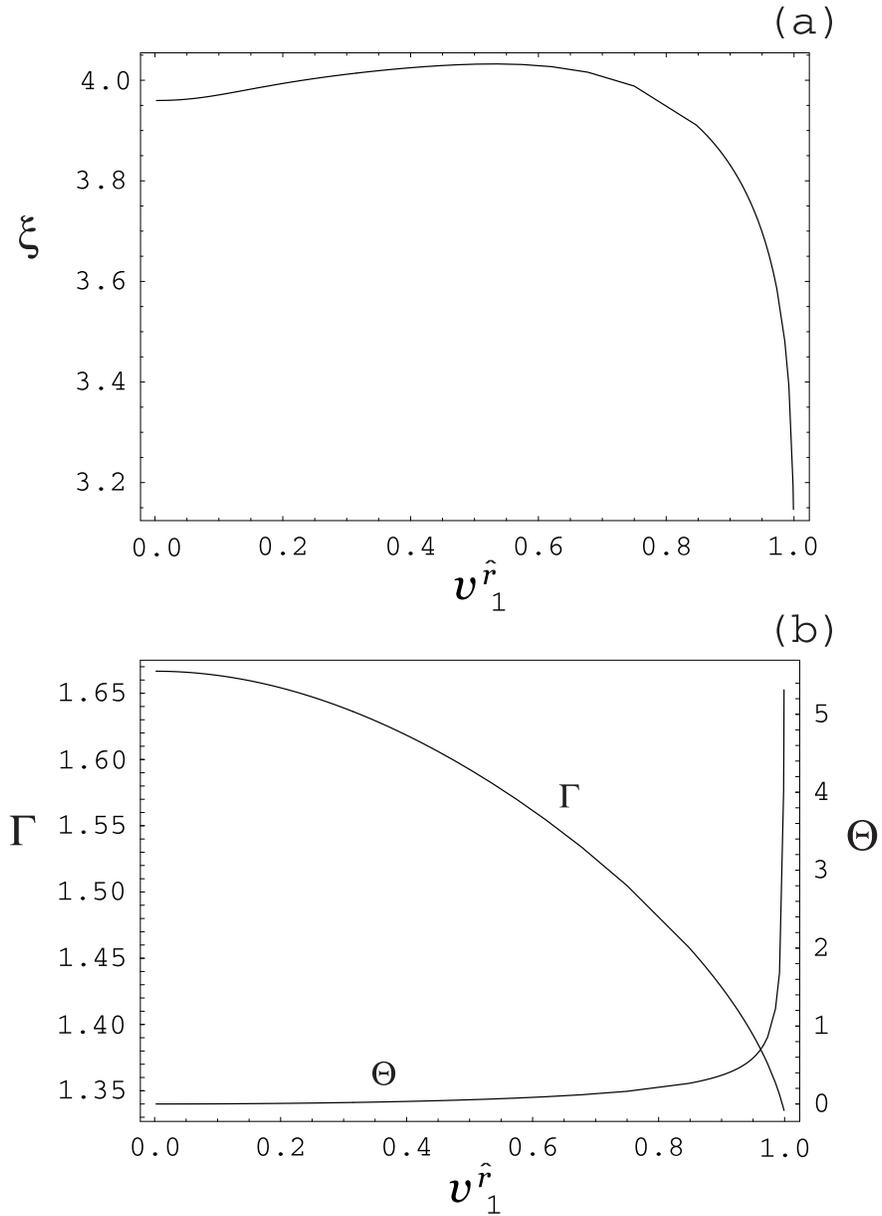}
    \caption{ 
 (a) The shock frame compression ratio $\xi$, and (b) polytropic index 
 $\Gamma$ and temperature parameter $\Theta$, respectively, as a  
 function of the radial preshock velocity $v^{\hat r}_1$ in the limit 
 of no plasma rotation and weak gravity ($\Omega_1=0$, $m=0$, $a=0$). 
 The set of field aligned parameters is given in  
 Table~\ref{tab:para-2}.  
 }
\label{fig:sk-AC}
\end{figure}

\placefigure{fig:sk-AC}

We extended the work by AC88 on the MHD shock conditions for special 
relativistic jets, by deriving corresponding equations for general 
relativistic accretion flows in the magnetosphere of a rotating black 
hole. Our results refer to the slow magnetosonic shock for accretion 
flows very close to a black hole where the general relativistic effects 
are  strong, while the AC88 results refer to the fast magnetosonic shock 
for outflows at long distances away from the hole. 
The one apparent difference between our results and those by AC88 is 
that their $r$ (our $\xi$) and $\Pi$ $\rightarrow$ 0 sharply as their 
$\beta$ $\rightarrow 0$, while our $\xi$ and $\Pi$ do not drop off 
sharply as $v^{\hat r}_1$ $\rightarrow$ 0 (see our 
Figures~\ref{fig:sk-1}a and~\ref{fig:sk-1}b). 
(In the rest of the velocity range, we find that the general trend of 
our results and theirs is quite similar, especially for positive $a$.)   
Another difference is the velocity dependence of $q$. 
For the fast magnetosonic shock in the case of AC88, there 
is a lower-limit threshold for the upstream velocity $v^{\hat r}_1$ 
to  satisfy the requirement of $q>1$. When this lower-limit velocity 
is approached (where $q=1$), the compression ratio decreases, and below 
this velocity there is no fast magnetosonic shock solutions. On the  
other hand, for the slow magnetosonic shock in our case, the condition 
$q<1$ is required, and moreover there is not such a threshold for 
the lower limit to the upstream velocity and $\xi$ does not drop off.  
For $v^{\hat r}_1 \to 0$ we obtain  $q=1$, and for the switch-off shock  
(if present) we obtain $q=0$ (see Fig.~\ref{fig:sk-2}a).  
The switch-off shock gives the upper limit to the upstream flow 
velocity, which is just the Alfv\'en wave velocity. 
Note that in the rotating black hole case the velocity range that 
allows to the slow magnetosonic shock is restricted by some other 
factors --- for instance, due to the dragging effect.

The general behavior of $\Gamma$ and $\Theta$ with velocity for all 
spin parameters $a$ as exhibited in Figures~\ref{fig:sk-1}c 
and~\ref{fig:sk-1}d are essentially the same as Figure~\ref{fig:sk-AC} 
of AC88 for hydrodynamic shocks with no plasma rotation. As mentioned 
by AC88, we also find that the polytropic index for the postshock 
plasma stays closer to $5/3$ than to $4/3$, except for extremely high 
Lorentz factors; here we should note that, for small $v^{\hat r}_1$,  
$\Gamma$ remains $\sim 5/3$ even when $v^{\hat \phi}_1$ dominates the 
flow velocity.

We note that our equations derived in Section~3.2 reduce to the 
corresponding equations in AC88 in the limit of no plasma rotation and  
weak gravity ($\Omega_1$ = 0, $m$ = 0 and $a$ = 0). With more direct 
comparison with AC88 in mind, we present in Figure~\ref{fig:sk-AC} 
the result for the asymptotic case of $a = 0$, 
$\Omega_1 = 0$, slowly rotating magnetosphere $\Omega_F\ll 1$,
and large distance, closer to the situation investigated by AC88. 
Figure~\ref{fig:sk-AC}a shows the compression ratio vs. $v^{\hat r}_1$ 
relation, while in Figure~\ref{fig:sk-AC}b the adiabatic exponent 
and temperature parameter are shown as a function of $v^{\hat r}_1$. 
The chosen conserved quantities are given in Table~\ref{tab:para-2}. 
These parameters appear to put us in the small $b^\prime$ and small 
$\theta^\prime$ regime (small $\sigma$) of 
AC88. Figures~\ref{fig:sk-AC}a and~\ref{fig:sk-AC}b show that although 
the parameters chosen here and AC88 are somewhat different, these 
curves are essentially identical to figures~1 and~6 of AC88.  
(Note that figure 6 in AC88 shows the hydrodynamical shock case.)

\begin{figure}
    \epsscale{0.9}
    \plotone{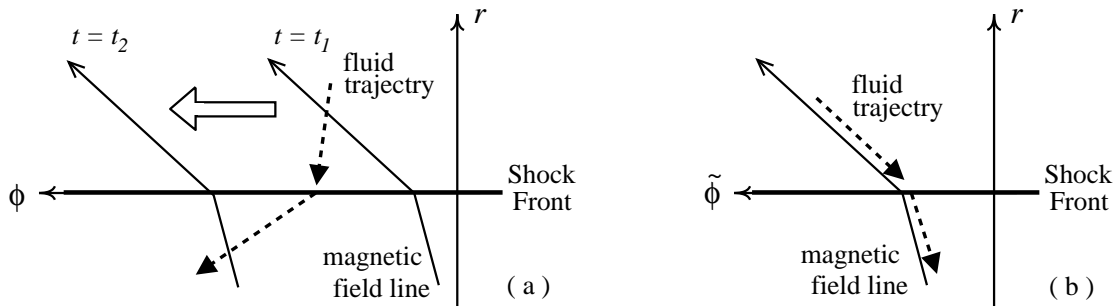}
    \caption{
 The slow magnetosonic shock seen by (a) a distant observer in the 
 ($r, \phi$)-plane and (b) a shock frame observer in the 
 ($r, \tilde \phi$)-plane, which is corotating with the magnetic field 
 line. For a slowly rotating black hole case ($\omega_{\rm H}<\Omega_F$), 
 the magnetic field lines are leading the black hole 
 toward the rotational direction, 
 and the fluid trajectory across the shock front looks like the fast 
 magnetosonic shock case; and then some physical properties of the slow 
 magnetosonic shock is very similar to that of the fast magnetosonic 
 shock. 
 }
\label{fig:slow-sk}
\end{figure}

Though our results of slow magnetosonic shocks for $a>0$ are very 
similar to the AC88 results obtained for the fast magnetosonic shocks, 
the local physical properties are obviously different between the 
fast and slow magnetosonic shocks. Here, we should note that our 
results include the rotational effects of magnetic field lines, which 
are leading the black hole toward the rotational direction. So this 
apparent similarity is caused essentially by the rotation of the 
magnetic field lines. 
When $0<\Omega_1<\Omega_F$, we find that $\Omega_1<\Omega_2<\Omega_F$.  
This means that in our slow magnetosonic shock case the trajectory of 
a fluid element seen by a distant observer becomes very similar to the 
fast magnetosonic shock case without rotation, as shown in 
Figure~\ref{fig:slow-sk}a. On the other hand, in the shock frame rotating 
with $\Omega_F$, the trajectory of a fluid element in our case is 
coincident with the magnetic field pattern (see Fig.~\ref{fig:slow-sk}b). 
So, some shock properties of the slow magnetosonic shock with 
$\Omega_1>0$ behave like the fast magnetosonic shock without rotation 
(especially the hydrodynamical behavior). 
 This is not essentially due to the general relativistic effects, 
but the existence of the slowly rotating black hole 
($0<\omega_{\rm H}<\Omega_F$), or a counter-rotating black hole 
($\Omega_F \omega_{\rm H}<0$), is responsible for such a situation 
(with the magnetic field lines leading the black hole in rotation) 
in the rotating magnetosphere. 
When $\Omega_1<0$, which state is easily realized for a rapidly 
counter-rotating black hole, the downstream plasma angular velocity 
$\Omega_2$ is in the range of $\Omega_1<\Omega_2\leq 0$ or 
$0<\Omega_2<\Omega_F$, and the trajectory of the fluid is quite 
different from the fast magnetosonic shock case, and then the shock 
behavior is also different. Also, for a rapidly rotating black hole 
with $0<\Omega_F<\omega_{\rm H}$ and the magnetic field lines trailing   
the black hole in the rotational direction, this similarity would 
be lost except for the very strong shock cases. Note that the dominant 
energy dissipation and accompanying radiation processes  
in the postshock hot plasma would be different between 
the {\it corotating}\/ slow magnetosonic shock and the fast magnetosonic 
shock without rotation, even if some of their properties are very similar.

\placetable{tab:para-2}

\placefigure{fig:sk-sigma}

Finally in Figure~\ref{fig:sk-sigma} we show our preliminary results 
for the upstream (solid curve) and downstream (dashed curve) 
magnetization parameters, $\sigma_1$ and $\sigma_2$, plotted as a 
function of the shock location, $r_{\rm sh}$, for a representative 
spin of $a=0.7m$. The conserved quantities chosen for this spin 
parameter are given in Table~\ref{tab:para-1}. From the energy momentum 
conservation at the shock front with equation (\ref{eq:Trt-c}), the 
postshock magnetization parameter $\sigma_2$ is expressed as 
$\sigma_2 = \zeta^{-1}\sigma_1 + (\zeta^{-1}-1)$. 
The decreasing $\sigma$ across the shock means that $\zeta > 1$. 
Although, at the shock front, kinetic energy of the preshock flow 
obviously converts to thermal energy of the postshock flow, the 
electromagnetic energy also converts to thermal energy at the same 
shock front. At the location of $r_{\rm sh} \to r_{\rm A}$ (when the 
switch-off shock takes place), the energy conversion is maximum. 
The detailed transitions of each of the energy components and the 
discussion of the implications will be presented in Paper II (Rilett 
et al., in preparation).

\section{Summary and Concluding Remarks}

The main purpose of this paper is to explore the general relativistic 
version of MHD shock conditions for accreting plasma in a magnetosphere 
of a rotating black hole.  Here, we introduce the magnetization 
parameter seen by ZAMO, because the shock conditions are related to 
the local plasma and the magnetosphere is dragged by the rotating black 
hole. Then, we can obtain similar expressions as for the special 
relativistic case, though the formulae include the Kerr metric. We have 
seen that the compression ratio $\xi$ is the solution of a polynomial 
of eighth degree.

Although the jump condition is of course determined by the local
physics, for a distant observer, the shock properties depend also 
on the gravitational field structure (i.e., the gravitational redshift 
factor  and the dragging effect, etc); furthermore, the preshock plasma  
flow parameters are also affected by the gravity and its spin.   
So, when we observe some physical properties of the  magnetosonic 
shock, which may be related to some efficient radiation process,  
we should understand the gravitational field structure in addition  
to the properties of magnetized plasma flow.

In this paper our sample solutions are presented for the special 
case of slow magnetosonic shocks in plasmas accreting in the 
equatorial plane with the radial magnetic field configuration. 
For slow magnetosonic shocks our results are compared with those by 
AC88. It is confirmed that, though AC88 discussed fast magnetosonic 
shocks, the general agreement is excellent in the limit of no plasma 
rotation and weak gravity. However, with the introduction of 
additional factors, gravity and rotation, we find that the situation 
becomes far more complicated. For example, in a black hole 
magnetosphere, two Alfv\'en points and two slow magnetosonic points 
can be located inside the separation point  $r=r_{\rm sp}$, where the 
gravitational force and the centrifugal force are balanced for plasma 
with zero poloidal velocity and the radius is defined by 
$(\alpha')_{\rm sp}=0$; the prime means the derivative along a stream 
line. In this paper we have treated only cold MHD accreting flow passing 
through the inner Alfv\'en point. However, the MHD accretion flow 
passing through the outer Alfv\'en point is also possible.  In both 
cases the injection point can be located near (inside) the separation 
point. When we consider shock formation where the postshock flow passes 
through the outer Alfv\'en point, we would find such a flow as an 
additional branch in the figures such as those presented in this paper. 
Such a new branch would appear outside the previous branch, and it 
of course must be located between the outer Alfv\'en point and the 
injection point.  However, the postshock accretion flow may pass 
through the inner Alfv\'en point. Furthermore, to obtain the accretion 
flow passing through the fast magnetosonic point twice by the shock 
formation, we must more carefully find the acceptable parameter 
ranges of conserved quantities. This will be our next step. 
In non-relativistic flow solutions, \cite{ch90b} discussed the 
slow and fast magnetosonic point conditions and shock conditions 
in winds, where multi-magnetosonic wind solutions were demonstrated. 
In the future, we will extend our results by treating a 
multi-magnetosonic flow conditions with shock formation in the Kerr 
geometry.

\begin{figure}
    \epsscale{0.7}
    \plotone{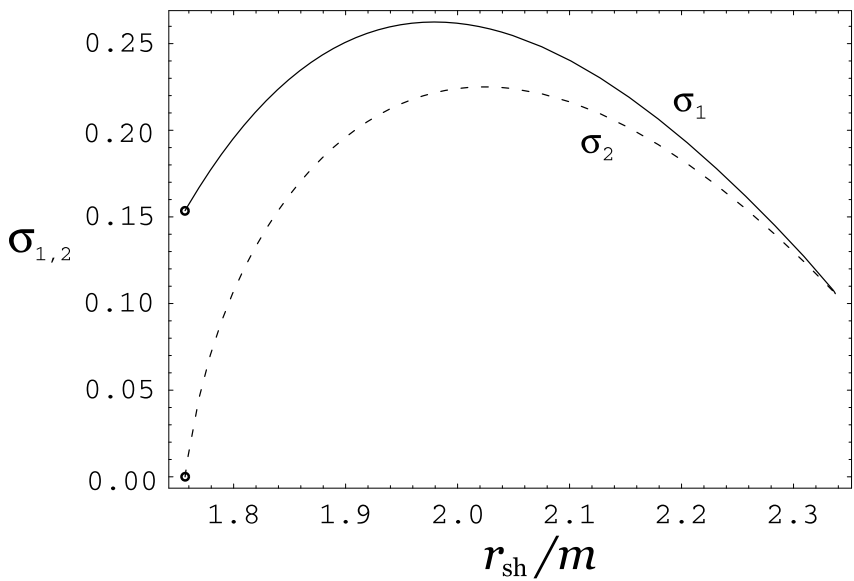}
    \caption{
 The upstream (solid curve) and downstream (dashed curve) magnetization
 parameters, $\sigma_1$ and $\sigma_2$, as a function of the shock 
 location, for $a=0.7m$ (see Table~\ref{tab:para-1} 
 for chosen flow parameters). 
 }
\label{fig:sk-sigma}
\end{figure}

As a preliminary example of how the transition of various parameters 
behave across the shock front, in Figure~\ref{fig:sk-sigma} we showed 
the preshock and postshock magnetization parameters. The detailed 
results for the transition behavior of various energy components 
and other parameters will be presented in Paper II (Rilett et al.,  
in preparation).  Here we only comment that both fluid kinetic  
energy and electromagnetic energy convert to thermal energy at the  
shock location.

When we consider accretion through high-latitude regions above 
the event horizon near the hole's rotational axis 
(see~\cite{ni91,tt01}), we can expect as an X-ray source hot 
ringed area around the axis.  Our extended future work will 
include the application of our current work to realistic models 
of a black hole magnetosphere as the central engine of 
accretion-powered AGN.

\acknowledgments

We are grateful to Akira Tomimatsu and Hideyo Kunieda for useful 
discussions. S.T. thanks the Yamada Foundation for financial 
support through the visiting program. 


\clearpage

\clearpage
%

\begin{table}
\begin{tabular}[t]{|c||l|l|}\hline
\begin{tabular}[c]{c}
Conserved  \\
Parameter  
\end{tabular} &  Physical description   &   Definition  \\ \hline\hline
$\Omega_F$ & the angular velocity of 
             the magnetic field lines    & ~~Eq.(8)   \\  \hline 
$E$        & the total energy 
             seen by a distant observer  & ~~Eq.(17)  \\  \hline 
$L$        & the total angular momentum 
             seen by a distant observer  & ~~Eq.(18)  \\  \hline 
$\eta$     & the particle number flux 
             per magnetic flux tube      & ~~Eq.(19)  \\  \hline 
\end{tabular}
\caption{
The table of conserved flow parameters. The mathematical definitions 
are expressed in the text (see, \cite{bk78,ca86}). 
}
\label{tab:add-1}
\end{table}

\begin{table}
\begin{tabular}[t]{|r||c|c||c|c|c|c|c|c|}\hline
$a/m$\ & $L/\mu_{1}$ & $m\Omega_F$ & $r_{\rm H}/m$ & $r_{\rm L}/m$ 
& $r_{\rm A}/m$ & $r_{\rm inj}/m$ & $r_{\rm sp}/m$         \\ \hline\hline
0.95 & 2.470 & 0.3691 & 1.312 & 1.314 & 1.335 & 1.439 & 1.457  \\  \hline 
0.70 & 3.137 & 0.2214 & 1.714 & 1.718 & 1.757 & 2.337 & 2.442  \\  \hline 
0.30 & 3.950 & 0.1096 & 1.954 & 1.964 & 2.022 & 3.694 & 4.270  \\  \hline 
0.00 & 5.073 & 0.0516 & 2.000 & 2.022 & 2.093 & 5.508 & 7.256  \\  \hline 
$-0.20$ & 7.332 & 0.0278 & 1.980 & 2.029 & 2.102 & 6.197 & 8.504  \\ \hline
$-0.40$ & 10.233 & 0.0062 & 1.916 & 2.010 & 2.090 & 21.406 & 29.609  \\ \hline
\end{tabular}
\caption{
 The parameter sets for cold trans-fast MHD accretion solutions, 
 where $E/\mu_{1} = 1.006$,  
 $\mu_{1}\eta = 0.041$ and $\mu_1=m_{\rm p}$ (for a cold flow). 
}
\label{tab:para-1}
\end{table}

\begin{table}
\begin{tabular}[t]{|c||c|c|c|c||c|c|c|}\hline
  $a/m$ & $E/\mu_{1}$     & $L/\mu_{1}$ & $m\Omega_F$ & $\mu_{1}\eta$
        & $r_{\rm L}/m$ & $r_{\rm A}/m$ & $r_{\rm inj}/m$
                                                       \\ \hline\hline
  0.0   & 1.00024  &  15.000    & 0.0019      & 0.0005
        & 2.00003  & 2.001      & 44.6                       \\ \hline
\end{tabular}
\caption{
 The parameter sets for cold trans-fast MHD accretion solutions.
}
\label{tab:para-2}
\end{table}

\end{document}